\documentclass[showkeys,showpacs,superscriptaddress]{revtex4}
\usepackage{amsmath}
\usepackage{amssymb}
\usepackage{graphicx}

\begin{document}

\title{Dirac star with SU(2) Yang-Mills and Proca fields
}
\author{Vladimir Dzhunushaliev}
\email{v.dzhunushaliev@gmail.com}
\affiliation{
	Institute of Experimental and Theoretical Physics,  Al-Farabi Kazakh National University, Almaty 050040, Kazakhstan
}
\affiliation{
National Nanotechnology Laboratory of Open Type,  Al-Farabi Kazakh National University, Almaty 050040, Kazakhstan
}
\affiliation{
	Department of Theoretical and Nuclear Physics,  Al-Farabi Kazakh National University, Almaty 050040, Kazakhstan
}
\affiliation{
Academician J.~Jeenbaev Institute of Physics of the NAS of the Kyrgyz Republic, 265 a, Chui Street, Bishkek 720071, Kyrgyzstan
}

\author{Vladimir Folomeev}
\email{vfolomeev@mail.ru}
\affiliation{
	Institute of Experimental and Theoretical Physics,  Al-Farabi Kazakh National University, Almaty 050040, Kazakhstan
}
\affiliation{
National Nanotechnology Laboratory of Open Type,  Al-Farabi Kazakh National University, Almaty 050040, Kazakhstan
}
\affiliation{
Academician J.~Jeenbaev Institute of Physics of the NAS of the Kyrgyz Republic, 265 a, Chui Street, Bishkek 720071, Kyrgyzstan
}


\begin{abstract}
We study spherically symmetric strongly gravitating configurations supported by nonlinear spinor fields and
non-Abelian SU(2) Yang-Mills/Proca magnetic fields. Regular asymptotically flat solutions describing objects
with positive Arnowitt-Deser-Misner masses are obtained numerically.
When the mass of the spinor fields is much smaller than the Planck mass, we find approximate solutions that can describe systems
with total masses comparable to the Chandrasekhar mass and with effective radii of the order of kilometers.
For the values of the system free parameters used here, we show that the SU(2) magnetic field always gives a small contribution to the
total energy density and mass of the configurations under investigation. From the astrophysical point of view, one can regard such objects as
magnetized Dirac stars.
\end{abstract}

\pacs{04.40.Dg, 04.40.--b, 04.40.Nr}

\keywords{Dirac star, nonlinear spinor fields, SU(2) Yang-Mills and Proca fields, compact gravitating configurations}

\maketitle

\section{Introduction}

Various aspects of constructing particlelike configurations supported by fundamental fields are widely studied in the literature.
Such configurations may consist of both integer-spin fields and spinor fields.
Spin-0 scalar fields have found their greatest use in astrophysical applications.
In the presence of gravitational fields, the scalar fields can create objects (the so-called boson stars)
whose physical characteristics can range from microscopic values up to those typical for
galaxies~\cite{Schunck:2003kk,Liebling:2012fv}.

On the other hand, nonzero-spin fields can also support the existence of particlelike objects.
In particular, these can be compact, strongly gravitating configurations supported by massless vector fields (Yang-Mills systems~\cite{Bartnik:1988am,Volkov:1998cc}) or
by massive vector fields (Proca stars~\cite{Brito:2015pxa,Herdeiro:2017fhv,Sanchis-Gual:2019ljs}). The fields of the latter type were initially used to model
short-range nuclear forces~\cite{Proca:1900nv}. Subsequently, such fields were applied, for instance, to describe the massive spin-1 $Z^0$ and $W^\pm$ bosons
in the standard model~\cite{Lawrie2002}, to account for various effects related to the possible presence of the rest mass of the photon~\cite{Tu:2005ge},
and within the framework of dark matter physics~\cite{Pospelov:2008jd}.

In turn, fractional-spin fields, being a source of gravitation, also allow for various compact objects.
If for such fields one uses spin-$1/2$ fields, it is possible to obtain spherically symmetric systems consisting of both linear
 \cite{Finster:1998ws,Finster:1998ux,Herdeiro:2017fhv} and nonlinear spinor fields~\cite{Krechet:2014nda,Adanhounme:2012cm,Dzhunushaliev:2018jhj,Bronnikov:2019nqa}.
Nonlinear spinor fields are also used to study cylindrically symmetric solutions~\cite{Bronnikov:2004uu}, wormhole solutions~\cite{Bronnikov:2009na},
and various cosmological problems (see Refs.~\cite{Ribas:2010zj,Ribas:2016ulz,Saha:2016cbu} and references inside).
We  also note the possibility of applying spinor fields in the modeling of compact objects, which consists in using some effective description of nonlinear spinor fields
in terms of complex scalar fields~\cite{Mielke:1980sa}.

In Ref.~\cite{Dzhunushaliev:2019kiy}, we considered a gravitating system consisting of two nonlinear Dirac fields
minimally coupled to U(1) Maxwell and Proca fields. For such a system, it was shown that for some values of the coupling constants one can obtain configurations
 with masses of the order of the Chandrasekhar mass. In the present paper, we extend those investigations and study the
effects associated with the presence of SU(2) Yang-Mills and Proca fields in the systems supported by nonlinear spinor fields.

It should be emphasized that in spite of the fact that realistic spin-$1/2$ particles must be described
by {\it quantum} spinor fields,  in the present paper we deal with {\it classical} spinor fields. By the latter we mean
a set of four complex-valued spacetime functions that transform according to
the spinor representation of the Lorentz group. Such classical spinors can arise from some effective description
of more complex quantum systems (for more details regarding this issue, see Ref.~\cite{ArmendarizPicon:2003qk}).
Also, one might think that a classical nonlinear spinor field can approximately
describe the interaction between sea quarks and gluons in QCD~\cite{Dzhunushaliev:2019ham}.

The paper is organized as follows. In Sec.~\ref{prob_statem}, we present the statement of the problem and write down
the general-relativistic equations for the systems under consideration.
These equations are solved numerically in Sec.~\ref{num_sol} for the Yang-Mills field
(Sec.~\ref{YM_case}) and the Proca field (Sec.~\ref{Proca_case}), including the limiting case of spinor particles with small masses (Sec.~\ref{YM_case_lim}).
Finally, in Sec.~\ref{concl} we summarize and discuss the results obtained.

\section{Statement of the problem and general equations}
\label{prob_statem}

We consider compact gravitating configurations consisting of a spinor field minimally coupled to non-Abelian SU(2)
 Yang-Mills/Proca fields. The modeling is carried out within the framework of Einstein's general relativity.
The corresponding total action for such a system can be represented in the form
[the metric signature is $(+,-,-,-)$]
\begin{equation}
\label{action_gen}
	S_{\text{tot}} = - \frac{c^3}{16\pi G}\int d^4 x
		\sqrt{-g} R +S_{\text{sp}}+S_{\text{v}},
\end{equation}
where $G$ is the Newtonian gravitational constant,
$R$ is the scalar curvature, and $S_{\text{sp}}$ and $S_{\text{v}}$ denote the actions
of spinor and vector fields, respectively.

The action $S_{\text{sp}}$ is obtained from the Lagrangian for the spinor field  $\psi$ of the mass $\mu$,
\begin{equation}
	L_{\text{sp}} =	\frac{i \hbar c}{2} \left(
			\bar \psi \gamma^\mu \psi_{; \mu} -
			\bar \psi_{; \mu} \gamma^\mu \psi
		\right) - \mu c^2 \bar \psi \psi - F(S),
\label{lagr_sp}
\end{equation}
where the semicolon denotes the covariant derivative defined as
$
\psi_{; \mu} =  [\partial_{ \mu} +1/8\, \omega_{a b \mu}\left( \gamma^a  \gamma^b- \gamma^b  \gamma^a\right)- i (g/2) \sigma^a
		A^a_\mu]\psi
$.
Here $\gamma^a$ are the Dirac matrices in the standard representation in flat space
 [see, e.g.,  Eq.~(7.27) of Ref.~\cite{Lawrie2002}]. In turn, the Dirac matrices in curved space,
$\gamma^\mu = e_a^{\phantom{a} \mu} \gamma^a$, are obtained using the tetrad
 $ e_a^{\phantom{a} \mu}$, and $\omega_{a b \mu}$ is the spin connection
[for its definition, see Eq.~(7.135) of Ref.~\cite{Lawrie2002}].
The term $i (g/2) \sigma^a A^a_\mu\psi$ describes the interaction between the spinor and Yang-Mills/Proca fields,
where $g$ is the coupling constant and $\sigma^a$ are the SU(2) generators (the Pauli matrices).
This Lagrangian contains an arbitrary nonlinear term $F(S)$, where the invariant $S$ can in general depend on
$
	\left( \bar\psi \psi \right),
	\left( \bar\psi \gamma^\mu \psi \right)
	\left( \bar\psi \gamma_\mu \psi \right)$, or
	$\left( \bar\psi \gamma^5 \gamma^\mu \psi \right)
	\left( \bar\psi \gamma^5 \gamma_\mu \psi \right)$.

The action for the vector fields $S_{\text{v}}$ appearing in Eq.~\eqref{action_gen} is obtained from the Lagrangian
$$	L_{\text{v}} =
		-\frac{1}{4} F^a_{\mu\nu}F^{a \mu\nu}+\frac{1}{2}\left(\frac{m_P c}{\hbar}\right)^2 A^a_\mu A^{a \mu},
$$
where $
	F^a_{\mu \nu} = \partial_\mu A^a_\nu - \partial_\nu A^a_\mu +
	g \epsilon_{a b c} A^b_\mu A^c_\nu
$ is the tensor of a massive spin-1 field of the Proca mass $m_P$;  $\epsilon_{a b c}$ (the completely antisymmetric Levi-Civita symbol)
are the SU(2) structure constants.
In the case of $m_P=0$ we return to a Yang-Mills system.

Varying the action \eqref{action_gen} with respect to the metric, the spinor field, and the vector potential $A_\mu$, we derive the Einstein, Dirac,
and Yang-Mills/Proca equations in curved spacetime:
\begin{eqnarray}
	R_{\mu}^\nu - \frac{1}{2} \delta_{\mu }^\nu R &=&
	\frac{8\pi G}{c^4} T_{\mu }^\nu,
\label{feqs-10} \\
	i \hbar \gamma^\mu \psi_{;\mu} - \mu c \psi - \frac{1}{c}\frac{\partial F}{\partial\bar\psi}&=& 0,
\label{feqs-20}\\
	i \hbar \bar\psi_{;\mu} \gamma^\mu + \mu c \bar\psi +
	\frac{1}{c}\frac{\partial F}{\partial\psi}&=& 0,
\label{feqs-21}\\
\frac{1}{\sqrt{-g}}\frac{\partial}{\partial x^\nu}\left(\sqrt{-g}F^{a\mu\nu}\right)+g \epsilon_{a b c} A^b_\nu F^{c\mu\nu}
&=&\frac{g\hbar c}{2}\bar\psi\gamma^\mu\sigma^a\psi
+\left(\frac{m_P c}{\hbar}\right)^2 A^{a\mu}.
\label{feqs-22}
\end{eqnarray}
The right-hand side of Eq.~\eqref{feqs-10} contains the energy-momentum tensor
 $T_{\mu}^\nu$, which can be represented (already in a symmetric form) as
\begin{align}
\label{EM}
\begin{split}
	T_{\mu}^\nu =&\frac{i\hbar c }{4}g^{\nu\rho}\left[\bar\psi \gamma_{\mu} \psi_{;\rho}+\bar\psi\gamma_\rho\psi_{;\mu}
-\bar\psi_{;\mu}\gamma_{\rho }\psi-\bar\psi_{;\rho}\gamma_\mu\psi
\right]-\delta_\mu^\nu L_{\text{sp}}
\\
&-F^{a\nu\rho}F^a_{\mu\rho}+\frac{1}{4}\delta_\mu^\nu F^a_{\alpha\beta}F^{a\alpha\beta}+
\left(\frac{m_P c}{\hbar}\right)^2\left(A^a_\mu A^{a\nu}-\frac{1}{2}\delta_\mu^\nu A^a_\rho A^{a\rho}\right).
\end{split}
\end{align}
Next, taking into account the Dirac equations \eqref{feqs-20} and \eqref{feqs-21}, the Lagrangian \eqref{lagr_sp} becomes
$$
	L_{\text{sp}} = - F(S) + \frac{1}{2} \left(
		\bar\psi\frac{\partial F}{\partial\bar\psi} +
		\frac{\partial F}{\partial\psi}\psi
	\right).
$$
For our purpose,  we choose the nonlinear term in a simple power-law form,
$F(S) = - k(k+1)^{-1}\lambda\left(\bar\psi\psi\right)^{k+1},
$
where $k$ and $\lambda$ are some free parameters. In what follows we set $k=1$ to give
\begin{equation}
	F(S) = - \frac{\lambda}{2} \left(\bar\psi\psi\right)^2.
\label{nonlin_term_2}
\end{equation}
The constant $\lambda$ appearing here has the following physical meaning:
the case of $\lambda > 0$ corresponds to attraction
and the case of $\lambda < 0$ to repulsion~\cite{Dzhunushaliev:2019kiy}.
In the absence of gravitation, classical spinor fields with this type of nonlinearity have been considered, for instance,
in Refs.~\cite{Finkelstein:1951zz,Finkelstein:1956,Soler:1970xp},
where it has been shown that the corresponding nonlinear Dirac equation has regular finite-energy solutions in a flat spacetime.
In turn, soliton-type solutions of the nonlinear Dirac equation in a curved background have been studied in Ref.~\cite{Mielke:2017nwt}
(see also references therein).

Since here we consider only spherically symmetric configurations, it is convenient to choose the spacetime metric in the form
\begin{equation}
	ds^2 = N(r) \sigma^2(r) (dx^0)^2 - \frac{dr^2}{N(r)} - r^2 \left(
		d \theta^2 + \sin^2 \theta d \varphi^2
	\right),
\label{metric}
\end{equation}
where $N(r)=1-2 G m(r)/(c^2 r)$, and the function $m(r)$ corresponds to the current mass of the configuration
enclosed by a sphere with circumferential radius $r$; $x^0=c t$ is the time coordinate.

In order to describe the spinor field, one must choose the corresponding  {\it Ansatz}  for $\psi$
compatible with the spherically symmetric line element \eqref{metric}.
Here, we use a stationary ansatz, which can be taken in the following form (see, e.g., Refs.~\cite{Soler:1970xp,Li:1982gf,Li:1985gf,Herdeiro:2017fhv}):
\begin{equation}
	\psi^T =2\, e^{-i \frac{E t}{\hbar}} \begin{Bmatrix}
		\begin{pmatrix}
			0 \\ - u \\
		\end{pmatrix},
		\begin{pmatrix}
			u \\ 0 \\
		\end{pmatrix},
		\begin{pmatrix}
			i v \sin \theta e^{- i \varphi} \\ - i v \cos \theta \\
		\end{pmatrix},
		\begin{pmatrix}
			- i v \cos \theta \\ - i v \sin \theta e^{i \varphi} \\
		\end{pmatrix}
	\end{Bmatrix},
\label{spinor}
\end{equation}
where $E/\hbar$ is the spinor frequency and
$v(r)$ and $u(r)$ are two real functions.
This  {\it Ansatz} ensures that the spacetime of the system under consideration remains static. Here, each row
 describes a spin-$1/2$ fermion, and these two fermions have the same mass $\mu$ and opposite spins.
Thus, the  {\it Ansatz}  \eqref{spinor} describes two Dirac fields whose
energy-momentum tensors are not spherically symmetric, but their sum gives a spherically symmetric energy-momentum tensor.
[Regarding the relationships between the {\it Ansatz} \eqref{spinor} and {\it Ans\"{a}tze} used in the literature, see Ref.~\cite{Dzhunushaliev:2018jhj}.]

For the Yang-Mills/Proca fields, we employ the standard SU(2) monopole  {\it Ansatz}
\begin{eqnarray}
	A^a_i &=&  \frac{1}{g} \left[ 1 - f(r) \right]
	\begin{pmatrix}
		 0 & \phantom{-}\sin \varphi &  \sin \theta \cos \theta \cos \varphi \\
		 0 & -\cos \varphi &  \sin \theta \cos \theta \sin \varphi \\
		 0 & 0 & - \sin^2 \theta
	\end{pmatrix} , \quad
	i = r, \theta, \varphi  \text{ (in polar coordinates)},
\label{2-20}\\
	A^a_t &=& 0 ,
\label{2-13}
\end{eqnarray}
describing a radial magnetic field.
Then, substituting the {\it Ans\"{a}tze} \eqref{spinor}-\eqref{2-13} and the metric  \eqref{metric} into the field equations
\eqref{feqs-10}, \eqref{feqs-20}, and  \eqref{feqs-22}, one can obtain the following set of equations:
\begin{eqnarray}
	&&\bar v^\prime + \left[
		\frac{N^\prime}{4 N} + \frac{\sigma^\prime}{2\sigma}+\frac{1}{x}\right] \bar v +\frac{f\bar v}{x\sqrt{N}}
+ \left[
		\frac{1}{\sqrt{N}} +8\bar \lambda\,\frac{\bar v^2 - \bar u^2}{\sqrt{N}}- \frac{\bar E}{\sigma N}
	\right]\bar u= 0,
\label{fieldeqs-1_dmls}\\
	&&\bar u^\prime + \left[
		\frac{N^\prime}{4 N} + \frac{\sigma^\prime}{2\sigma}+\frac{1}{x}\right] \bar u -\frac{f\bar u}{x\sqrt{N}}
+ \left[
		\frac{1}{\sqrt{N}} +8\bar \lambda\,\frac{\bar v^2 - \bar u^2}{\sqrt{N}}+ \frac{\bar E}{\sigma N}
	\right]\bar v= 0,
\label{fieldeqs-2_dmls}\\
	 &&\bar m^\prime= \alpha x^2\left[2\frac{\left(f^2-1\right)^2+2 x^2 N f^{\prime 2}}{x^4}+
\frac{8\bar E}{\sigma\sqrt{N}}\left(\bar u^2+\bar v^2\right)+32\bar \lambda\left(\bar v^2-\bar u^2\right)^2
+4\beta^2\frac{(f-1)^2}{x^2}
\right],
\label{fieldeqs-3_dmls}\\
&&\frac{\sigma^\prime}{\sigma}	=8\alpha\frac{x}{\sqrt{N}}\left[
\frac{\bar E}{\sigma N}\left(\bar u^2+\bar v^2\right)+ \bar u \bar v^\prime-\bar v \bar u^\prime+\frac{\sqrt{N}f^{\prime 2}}{x^2}
\right],
\label{fieldeqs-4_dmls}\\
&& f^{\prime\prime}+\left(\frac{N^\prime}{N}+\frac{\sigma^\prime}{\sigma}\right)f^\prime+\frac{f}{x^2 N}\left(1-f^2\right)=2x\frac{\bar u\bar v}{N}
+\beta^2\frac{f-1}{N},
\label{fieldeqs-5_dmls}
\end{eqnarray}
where the prime denotes differentiation with respect to the radial coordinate.
Here,  Eqs.~\eqref{fieldeqs-3_dmls} and \eqref{fieldeqs-4_dmls} are the  ($^0_0$) and  $[(^0_0)~-~(^1_1)]$ components of the Einstein equations,
respectively. The above equations are written in terms of the following dimensionless variables and parameters:
\begin{eqnarray}
\begin{split}
\label{dmls_var}
	x = & r/\lambda_c, \quad
	\bar E = \frac{E}{\mu c^2}, \quad
	\bar v, \bar u = \sqrt{\frac{4\pi}{\alpha}}\lambda_c^{3/2}\frac{\mu}{M_\text{Pl}} v, u=2\lambda_c^{3/2} \bar g \, v, u,\quad
	\bar m = \frac{\mu}{M_\text{Pl}^2} m,
\\
	\bar \lambda = & \frac{\alpha}{4\pi \lambda_c^3\mu c^2}
	\left( M_\text{Pl}/\mu\right)^2\lambda=\frac{\lambda}{4\bar g^2 \lambda_c^3\mu c^2},  \quad
   \bar g^2=g^2\hbar c, \quad \alpha=\frac{\pi}{\bar g^2} \left( \mu/M_\text{Pl}\right)^2,
   \quad \beta=m_P/\mu,
\end{split}
\end{eqnarray}
where $M_\text{Pl}$ is the Planck mass and $\lambda_c=\hbar/\mu c$ is a constant with
the dimensions of length (since we consider a classical theory, $\lambda_c$ need not be associated with the Compton length);
 the metric function $N=1-2\bar m/x$. Notice here that, using the Dirac equations \eqref{fieldeqs-1_dmls} and \eqref{fieldeqs-2_dmls},
 one can eliminate the derivatives of $\bar v$ and $\bar u$ from the right-hand side of Eq.~\eqref{fieldeqs-4_dmls}.

In turn, consistent with the dimensions of  $[\lambda]=\text{erg cm}^3$, one can assume that its characteristic value is
  $\lambda \sim  \tilde \lambda \,\mu c^2 \lambda_c^3$, where the dimensionless quantity $\tilde \lambda \sim 1$.
Then, one can obtain from Eq.~\eqref{dmls_var} that
$
\bar \lambda=\tilde \lambda/\left(4\bar g^2\right).
$
By assuming that on the energy scales considered here
the coupling constant $\bar g$ is of the order of unity, the dimensionless parameter $\bar \lambda$ should also be of the order of unity.

However, here we should make the following remark. As is known from quantum chromodynamics~\cite{Lawrie2002},
 the coupling constant $g$ is in general energy dependent: the greater the energy, the smaller the magnitude of $g$.
 Correspondingly, this may affect the characteristics of the systems studied in the present paper. Namely,
one might expect that more massive configurations
will possess a greater characteristic energy density, and for them the value of $g$ will be smaller, with the result
that $\bar \lambda$ will increase as $1/\bar g^{2}$.
In principle, this must in general lead to a change of the physical properties of the compact configurations under investigation
(see, however, Sec.~\ref{YM_case_lim} where we  consider configurations whose masses and sizes do not depend on the magnitude of $g$).

\section{Numerical results}
\label{num_sol}

In performing the numerical integration of Eqs.~\eqref{fieldeqs-1_dmls}-\eqref{fieldeqs-5_dmls},
we start from the center of the configuration where some values of the spinor field $\bar u_c$  and the metric function $\sigma_c$ are given.
The boundary conditions in the vicinity of the center are taken in the form
\begin{equation}
	\bar u\approx \bar u_c + \frac{1}{2}\bar u_2 x^2, \quad
	\bar v\approx \bar v_1 x,
	\quad \sigma\approx \sigma_c+\frac{1}{2}\sigma_2 x^2, \quad
	\bar m\approx \frac{1}{6}\bar m_3 x^3, \quad f \approx 1 + \frac{1}{2} f_2 x^2.
\label{bound_cond}
\end{equation}
Expressions for the expansion coefficients $\bar v_1,  \bar m_3, \sigma_2$, and $\bar u_2$ can be found from
Eqs.~\eqref{fieldeqs-1_dmls}-\eqref{fieldeqs-5_dmls}. In turn, the expansion coefficients
$\sigma_c$, $\bar u_c$, and $f_2$, and also the parameter $\bar E$, are arbitrary.
Their values are chosen so as to obtain regular and asymptotically flat solutions when
the functions $N(x\to \infty),\sigma(x\to \infty) \to 1$, and $f(x\to \infty) \to \pm 1$. 
In this case, the asymptotic value of the function $\bar m(x\to \infty) \equiv \bar m_\infty= M/\left(M_\text{Pl}^2/\mu\right)$
will correspond to the total Arnowitt-Deser-Misner (ADM) mass of the configurations under consideration.

Since the spinor fields decrease exponentially with distance as $\bar u, \bar v \sim e^{-\sqrt{1-\bar E^2}\,x}$ [see Eqs.~\eqref{asymp_YM} and \eqref{asymp_Proca} below],
numerical calculations are performed up to some boundary point $x_b$
where both the functions $\bar u$ and $\bar v $ and their derivatives go to zero (let us refer to such solutions as interior ones).
The location of the boundary point is determined by both the central values $\bar u_c$ and $\sigma_c$
and the magnitudes of the parameters $\bar \lambda$, $\alpha$, and $\beta$. The interior solutions are smoothly matched with the exterior ones for
 $f$, the mass function $\bar m$, and the metric function  $\sigma$ on the boundary $x=x_b$. 
In doing so, we must choose such eigenvalues of $f_2$ and $\bar E$
for which regular monotonically damped solutions do exist.

\subsection{The case of the Yang-Mills field}
\label{YM_case}

Consider first the case of the Yang-Mills field, i.e., the problem with $\beta=0$. In our previous papers~\cite{Dzhunushaliev:2018jhj,Dzhunushaliev:2019kiy},
the major emphasis was placed on obtaining starlike configurations with total masses of the order of the Chandrasekhar mass.
It was shown there that it is possible to get such masses only for the limiting configurations with negative $\bar \lambda$, which have $|\bar \lambda|/\alpha \gg 1$
(in terms of the dimensionless variables used here). As it will be shown below, a similar situation takes place for the systems considered in the present paper.
Taking into account that we assume that $\bar \lambda\sim 1$ (see the end of Sec.~\ref{prob_statem}),
this corresponds to the fact that to obtain masses of the order of the Chandrasekhar mass, it is necessary to choose $\alpha\to 0$.
In turn, for $\alpha \sim 1$, the characteristics of the systems will be Planckian ones.

\begin{figure}[t]
	\begin{minipage}[t]{.49\linewidth}
		\begin{center}
			\includegraphics[width=1\linewidth]{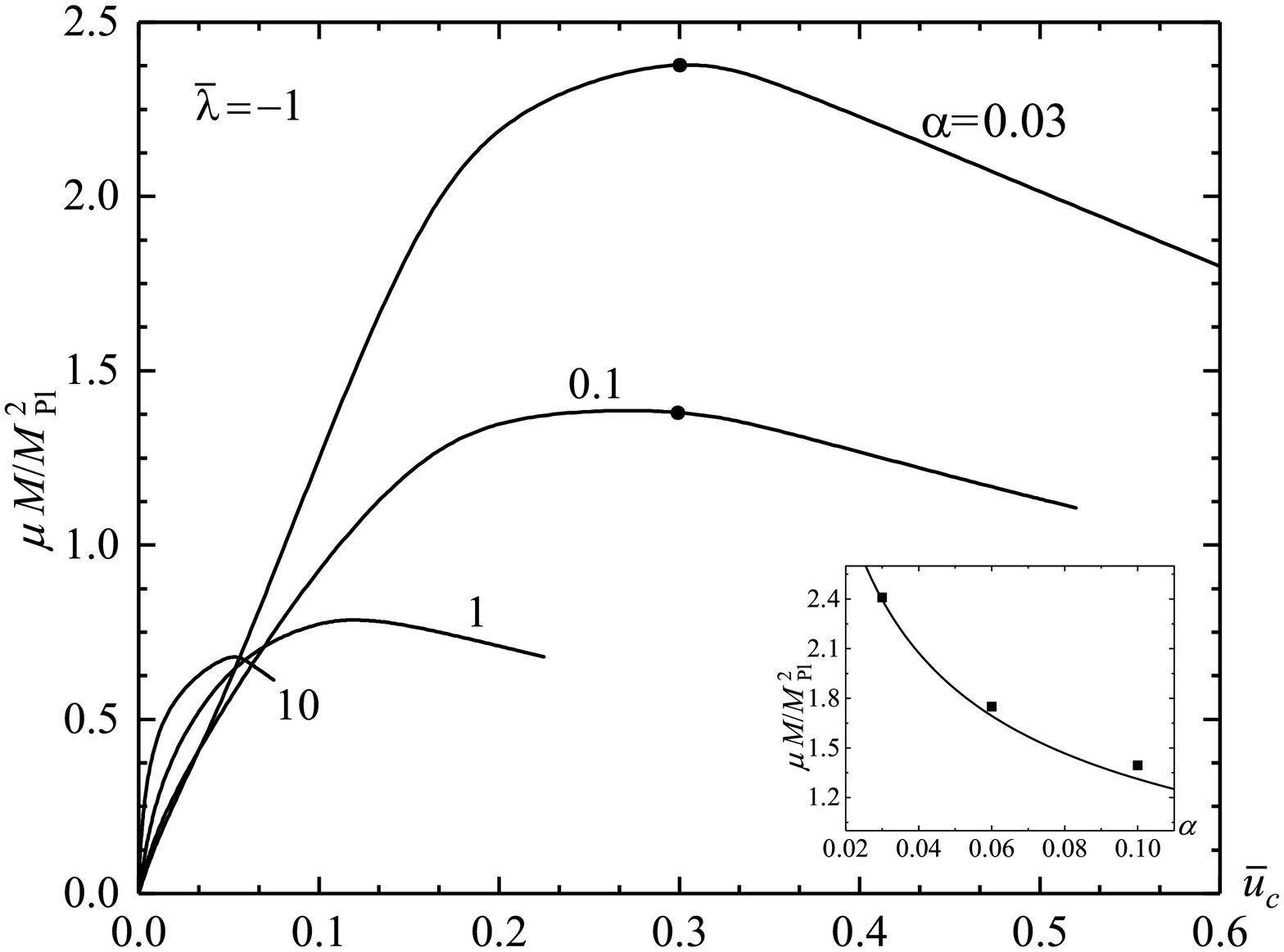}
		\end{center}
\vspace{-0.5cm}
		\caption{
		Dimensionless total ADM mass $\bar m_\infty$ as a function of $\bar u_c$ for $\alpha=0.03, 0.1, 1,$ and $10$.
The bold dots mark the positions of the configurations for which the graphs of Fig.~\ref{fig_field_distr} are plotted.
The inset depicts maximum masses for different $\alpha$ (shown by the square symbols);
the solid curve corresponds to the asymptotic relation~\eqref{M_max_approx}.
		}
		\label{fig_mass_uc}
	\end{minipage}\hfill
\begin{minipage}[t]{.49\linewidth}
		\begin{center}
			\includegraphics[width=.93\linewidth]{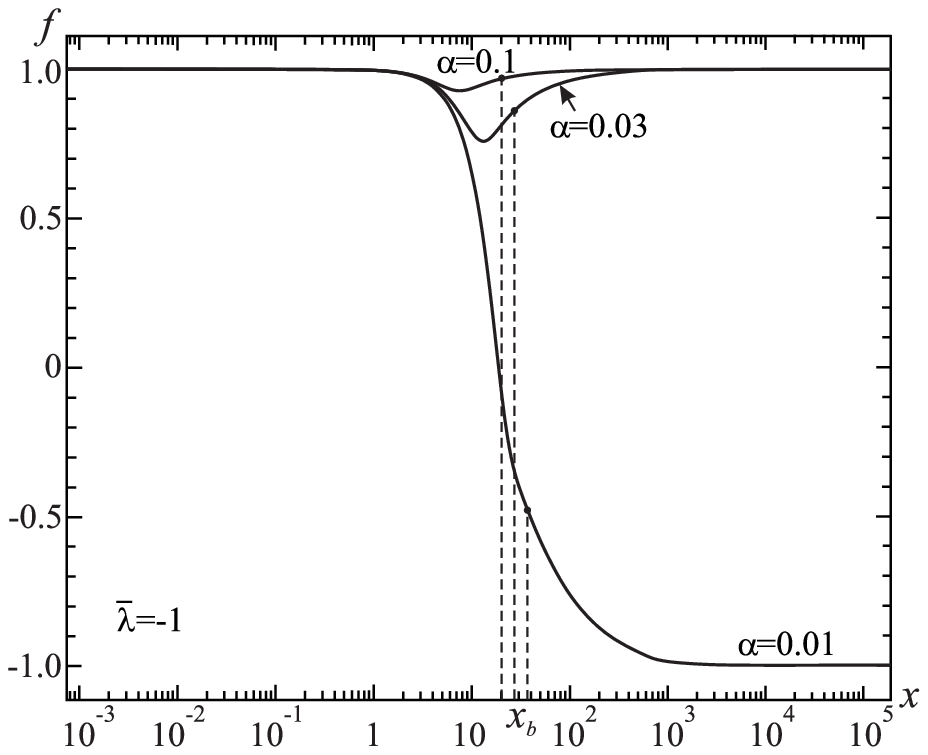}
		\end{center}
\vspace{-0.5cm}
		\caption{
		Yang-Mills field $f$ as a function of dimensionless radius~$x$ for $\alpha=0.01, 0.03$, and $0.1$ and $\bar u_c=0.15$.
The dashed lines show the positions of the boundary points $x_b$.
 		}
		\label{fig_f_alpha}
	\end{minipage}
\end{figure}

Taking all of this into account,
we set $\bar \lambda =-1$ and, by choosing different values of the parameter $\alpha$,
compute the total ADM mass of the system as a function of the central value of the spinor field $\bar u_c$. The results of numerical calculations are shown in Fig.~\ref{fig_mass_uc}.
Notice that for $\bar \lambda \geq 0$ one obtains configurations possessing considerably smaller masses
(cf. Refs.~\cite{Dzhunushaliev:2018jhj,Dzhunushaliev:2019kiy});
since our main purpose is to get configurations with masses of the order of the Chandrasekhar mass,
we do not show here the results of calculations for such values of $\bar \lambda$.

In plotting the dependencies of Fig.~\ref{fig_mass_uc}, we have kept track of the sign of the binding energy (BE),
which is defined as the difference between the energy of $N_f$ free particles, ${\cal E}_f=N_f \mu c^2$, and the total energy of the system, ${\cal E}_t=M c^2$,
i.e., $\text{BE}={\cal E}_f-{\cal E}_t$. Here, the total particle number
 $N_f$ (the Noether charge) is calculated using the timelike component of the 4-current $j^\alpha=\sqrt{-g}\bar \psi \gamma^\alpha \psi$
as
$
N_f=\int j^t d^3 x,
$
where in our case $j^t = N^{-1/2}r^2 \sin{\theta} \left(\psi^\dag \psi\right)$.
In the dimensionless variables \eqref{dmls_var}, we then have
\begin{equation}
N_f=8\alpha\left(\frac{M_\text{Pl}}{\mu}\right)^2\int_0^\infty \frac{\bar u^2+\bar v^2}{\sqrt{N}}x^2 dx.
 \label{part_num}
\end{equation}
A necessary condition for the energy stability is the positiveness of the BE.
Therefore, since configurations with a negative BE are certainly unstable, the graphs in Fig.~\ref{fig_mass_uc}
are plotted only up to $\bar u_c$ for which the BE becomes equal to zero (the rightmost points of the curves).

The typical distributions of the spinor and magnetic fields along the radius of the configurations are shown in Figs.~\ref{fig_f_alpha} and \ref{fig_field_distr}.
At the point $x=x_b$ the interior solutions for the functions $\bar m, \sigma$, and $f$ are smoothly matched with the exterior solutions describing the systems
for $x>x_b$. Such exterior solutions are obtained from the set of equations
\eqref{fieldeqs-3_dmls}-\eqref{fieldeqs-5_dmls}, where one sets $\bar u, \bar v=0$. The examples of the exterior solutions for the Yang-Mills field are given in Fig.~\ref{fig_f_alpha}.
Asymptotically (as $x\to \infty$), the solutions behave as
\begin{equation}
\bar v\approx \bar v_\infty e^{-\sqrt{1-\bar E^2}\,x}, \quad \bar u\approx \bar u_\infty e^{-\sqrt{1-\bar E^2}\,x}, \quad
f\approx \pm \left(1 - \frac{f_\infty}{x}\right),
\quad \sigma\approx 1-\frac{2\alpha f_\infty^2}{x^4}, \quad \bar m\approx \bar m_\infty-\frac{4\alpha  f_\infty^2}{x^3},
\label{asymp_YM}
\end{equation}
where $\bar v_\infty, \bar u_\infty,  f_\infty$, and $\bar m_\infty$ are integration constants, and the constant
$\bar m_\infty$ plays the role of the total ADM mass of the configurations under consideration.

The results of numerical calculations indicate the following.
\begin{itemize}
  \item The SU(2) magnetic field has a weak influence on the general structure of the configurations.
For this reason,  the behavior of the spinor fields
of the systems considered here practically coincides with that of the fields of the Dirac stars studied in Ref.~\cite{Dzhunushaliev:2018jhj}.
\item The behavior of the function $f$, which determines the distribution of the magnetic field,
depends strongly on the value of the parameter $\alpha$. For the magnitudes of $\alpha$ which are appreciably different from zero, the function  $f$,
starting from the origin of coordinates from unity, asymptotically goes to unity again. For $\alpha \ll 1$, the behavior of $f$ changes drastically: starting from unity
at the center, asymptotically $f\to -1$ (see Fig.~\ref{fig_f_alpha}).
That is, there exists some critical $\alpha_{\text{crit}}$ separating two types of solutions (in the terminology of Ref.~\cite{Rajaraman:1982is}):
topological (kinklike) solutions for $\alpha<\alpha_{\text{crit}}$ and nontopological solutions for $\alpha>\alpha_{\text{crit}}$. The behavior of $f$ for $\alpha<\alpha_{\text{crit}}$
is similar to the behavior of the magnetic field found by Bartnik and McKinnon~\cite{Bartnik:1988am,Volkov:1998cc}:
the solutions start from one minimum of a potential and tend asymptotically to another one.
\item Unlike the Bartnik-McKinnon solutions, when a solution for the Yang-Mills field has at least one node~\cite{Volkov:1998cc},
for the Dirac-Yang-Mills system considered here, there also exist nodeless solutions,
which are exemplified in Fig.~\ref{fig_f_alpha} by the curves for $\alpha=0.03$ and $\alpha=0.1$.
\end{itemize}

It is seen from Fig.~\ref{fig_mass_uc} that for all $\alpha$ there are maxima of the mass at some values of
$\bar u_c$. Such a behavior of the curves resembles the behavior of the corresponding
mass--central density dependencies for boson stars supported by a complex scalar field
(see, e.g., Refs.~\cite{Colpi:1986ye,Gleiser:1988ih,Herdeiro:2017fhv}). In the case of boson stars, such a maximum corresponds to the point
separating configurations that are stable or unstable against linear perturbations~\cite{Gleiser:1988ih}. One might expect that for the systems considered here a similar situation will also take place.
But this issue requires special study.

\subsection{Limiting configurations as $\alpha\to 0$}
\label{YM_case_lim}

\begin{figure}[t]
	\begin{minipage}[t]{.49\linewidth}
		\begin{center}
			\includegraphics[width=1\linewidth]{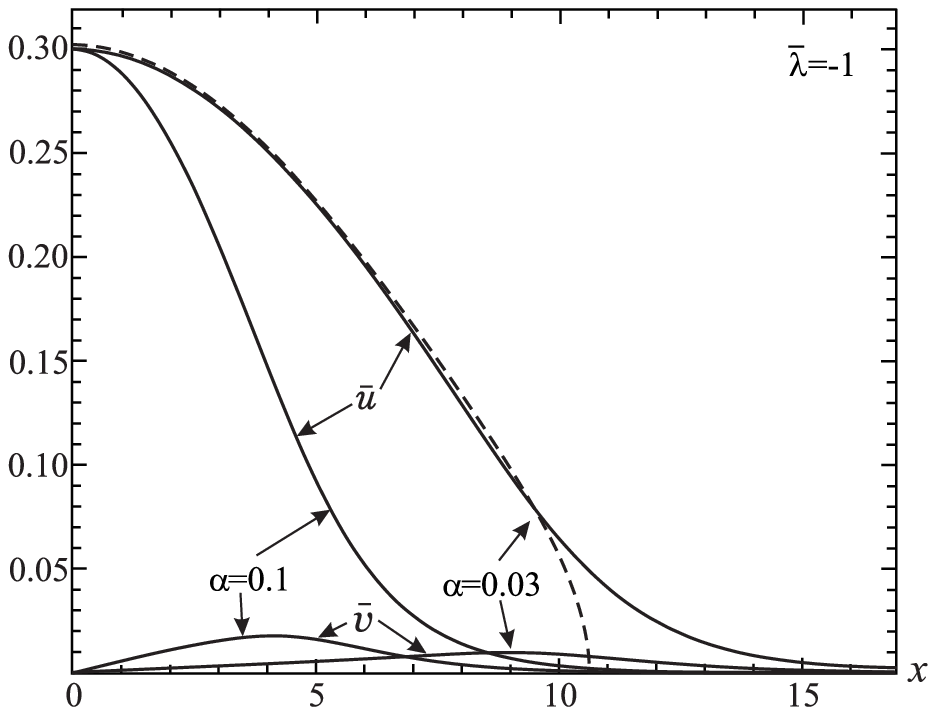}
		\end{center}
\vspace{-0.5cm}
		\caption{
		Spinor fields $\bar u$ and $\bar v$  as functions of dimensionless radius $x$ for $\alpha=0.03$ and $\alpha=0.1$.
The dashed line shows the solution to Eqs.~\eqref{u_approx}-\eqref{fieldeqs-4_dmls_approx}  with $\bar E/\sigma_c$
from the exact $\bar u_c=0.3, \alpha=0.03$ model, scaled to $\alpha=0.03$.
		}
		\label{fig_field_distr}
	\end{minipage}\hfill
\begin{minipage}[t]{.49\linewidth}
		\begin{center}
			\includegraphics[width=.95\linewidth]{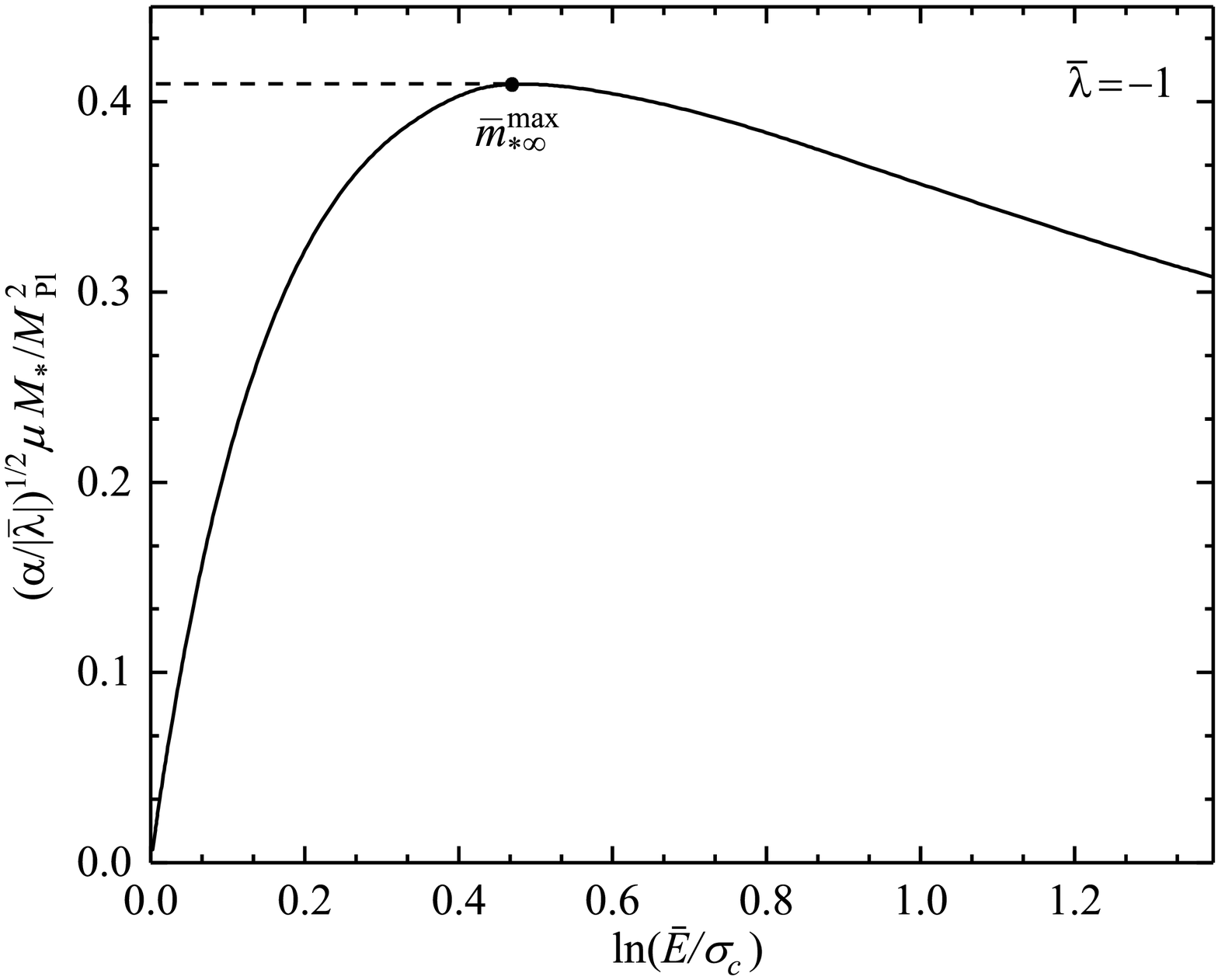}
		\end{center}
\vspace{-0.5cm}
		\caption{
		Dimensionless total mass $\bar m_{*\infty}$ as a function of $\bar E/\sigma_c$ for the limiting configurations described by Eqs.~\eqref{u_approx}-\eqref{fieldeqs-4_dmls_approx}.
The graph is plotted only for the values of $\bar E$ for which the binding energy is positive.
 		}
		\label{fig_mass_E_sigma}
	\end{minipage}
\end{figure}

It is seen from Fig.~\ref{fig_mass_uc} that when $\alpha$  decreases, the  maxima of the total mass of the configurations $M^{\text{max}}$ increase.
For better clarity, the inset of Fig.~\ref{fig_mass_uc} shows the dependence of
 $M^{\text{max}}$ on $\alpha$ for small $\alpha$. In this inset, the solid line corresponds to the interpolation formula
 \begin{equation}
\label{M_max_approx}
	M^{\text{max}}\approx \frac{0.415}{\sqrt{\alpha}}\frac{M_\text{Pl}^2}{\mu},
\end{equation}
which holds asymptotically for $\alpha \to 0$.

In order to study the solutions for small $\alpha$, we use the fact that, according to the results of numerical calculations, as in the case of the systems of
 Refs.~\cite{Dzhunushaliev:2018jhj,Dzhunushaliev:2019kiy}, there are some limiting solutions when for $\alpha\to 0$ the function $\bar v$
 becomes much smaller than $\bar u$. In this case one can obtain from Eq.~\eqref{fieldeqs-1_dmls} the approximate algebraic expression
\begin{equation}
 \label{u_approx}
 \bar u_* = \sqrt{-\frac{1}{8}
 \left(1 - \frac{\bar E}{\sigma\sqrt{N}}\right)},
\end{equation}
where we have introduced a new variable $\bar u_*=\sqrt{|\bar \lambda|}\bar u$.
(For more details regarding the validity of this approximation, see Refs.~\cite{Dzhunushaliev:2018jhj,Dzhunushaliev:2019kiy}.)
Substituting this expression into Eqs.~\eqref{fieldeqs-3_dmls} and \eqref{fieldeqs-4_dmls} and introducing the new variables
$x_*=\sqrt{\alpha/|\bar \lambda|} x$ and $\bar m_*=\sqrt{\alpha/|\bar \lambda|} \bar m$
(this corresponds to the scale invariance of these equations),
one can obtain the set of approximate equations
\begin{eqnarray}
	\frac{d \bar m_*}{d x_*} &=& 8 x_*^2 \left(
\frac{\bar E}{\sigma\sqrt{N}}-4 \bar u_*^2
	\right)\bar u_*^2,
\label{fieldeqs-3_dmls_approx}\\
\frac{d \sigma}{d x_*}& =&8 \frac{ x_*\sigma}{N}\left(\frac{2\bar E}{\sigma\sqrt{N}}-8 \bar u_*^2-1\right)\bar u_*^2,
\label{fieldeqs-4_dmls_approx}
\end{eqnarray}
where now $N=1-2\bar m_*/x_*$. As  $\alpha\to 0$, the accuracy of
Eqs.~\eqref{u_approx}-\eqref{fieldeqs-4_dmls_approx} becomes better.
This is illustrated in Fig.~\ref{fig_field_distr} where the results of calculations for the configurations with the same
$\bar u_{c}$ and for $\alpha=0.1$ and  $0.03$ are shown.
From comparison of the exact and approximate solutions, one can see their good agreement already for the case of $\alpha=0.03$,
except for the behavior at large radii.
As  $\alpha\to 0$, this region becomes less important and, accordingly, the mass of the configurations will be well described by the asymptotic formula.

Since
$\alpha$ does not appear explicitly in Eqs.~\eqref{fieldeqs-3_dmls_approx} and \eqref{fieldeqs-4_dmls_approx}, one can use these limiting equations to determine the rescaled total mass
$\bar m_{*\infty}\equiv\bar m_*(x\to\infty) = \sqrt{\alpha/|\bar \lambda|}M_{*}/\left(M_\text{Pl}^2/\mu\right)$
as a function of the single free parameter $\bar E/\sigma_c$.
The corresponding results of the numerical solution of Eqs.~\eqref{u_approx}-\eqref{fieldeqs-4_dmls_approx} are given in Fig.~\ref{fig_mass_E_sigma},
from which one can see the presence of a maximum of the mass,
\begin{equation}
\label{M_max_approx_2}
	M_*^{\text{max}} \approx 0.41\sqrt{\frac{|\bar \lambda|}{\alpha}}  \frac{M_\text{Pl}^2}{\mu}.
\end{equation}
Letting $\bar \lambda=-1$, it is seen that this expression agrees very well with that given in Eq.~\eqref{M_max_approx},
and this confirms that the above approximation is in good
agreement with the exact solution (cf. the results of Refs.~\cite{Dzhunushaliev:2018jhj,Dzhunushaliev:2019kiy}).

Here two points should be noted. First,
the numerical computations indicate that the above approximate solutions describe fairly well only systems located near maxima of the total mass,
and the deviations from the exact solutions become stronger the further away we are from the maximum.
Second, taking into account that the parameters $\bar \lambda$ and $\alpha$ appearing in
Eq.~\eqref{M_max_approx_2} are in inverse proportion to the square of the coupling constant $\bar g$
[see Eq.~\eqref{dmls_var}], the total mass of the limiting configurations, as well as their effective radius
[see Eq.~\eqref{R_eff_alpha_small}], do not depend on $\bar g$ (in this connection see also
the discussion given at the end of Sec.~\ref{prob_statem}).

\subsection{Effective radius}
\label{eff_rad_YM}

Let us now compute the radius of the configurations under consideration. Unlike ordinary stars (for example, neutron stars),
when it is assumed that there is a surface on which the pressure of matter vanishes, in the case of field configurations such a surface is already absent.
Therefore, in modeling such configurations, one uses some effective radius.
Since the configurations considered above contain a long-range Yang-Mills field, a definition of the effective radius
can be introduced by analogy with the case of charged boson stars with a Maxwell field in the form~\cite{Jetzer:1989av}
\begin{equation}
R=\frac{1}{N_f}\int r j^t d^3 x=\frac{\lambda_c}{N_f}\int_0^\infty \frac{\bar u^2+\bar v^2}{\sqrt{N}}x^3 dx,
\label{eff_radius}
\end{equation}
where the particle number $N_f$ is taken from Eq.~\eqref{part_num} (without the numerical coefficient before the integral).
This expression yields a finite result, in contrast to the case when the effective radius is defined in terms
of the mass integral, as is done for uncharged configurations (for details see Refs.~\cite{Schunck:2003kk,Jetzer:1989av}).

Using the expression~\eqref{eff_radius}, we have obtained the following dependence of the effective radius on  $\alpha$
for the limiting configurations with maximum masses considered in Sec.~\ref{YM_case_lim}:
\begin{equation}
 R_*^{\text{max}}\approx 1.08 \sqrt{\frac{|\bar \lambda|}{\alpha}}\lambda_c  \quad \text{for} \quad \alpha\ll 1.
\label{R_eff_alpha_small}
\end{equation}

\subsection{The case of the Proca field}
\label{Proca_case}

\begin{figure}[t]
	\begin{minipage}[t]{.49\linewidth}
		\begin{center}
			\includegraphics[width=1\linewidth]{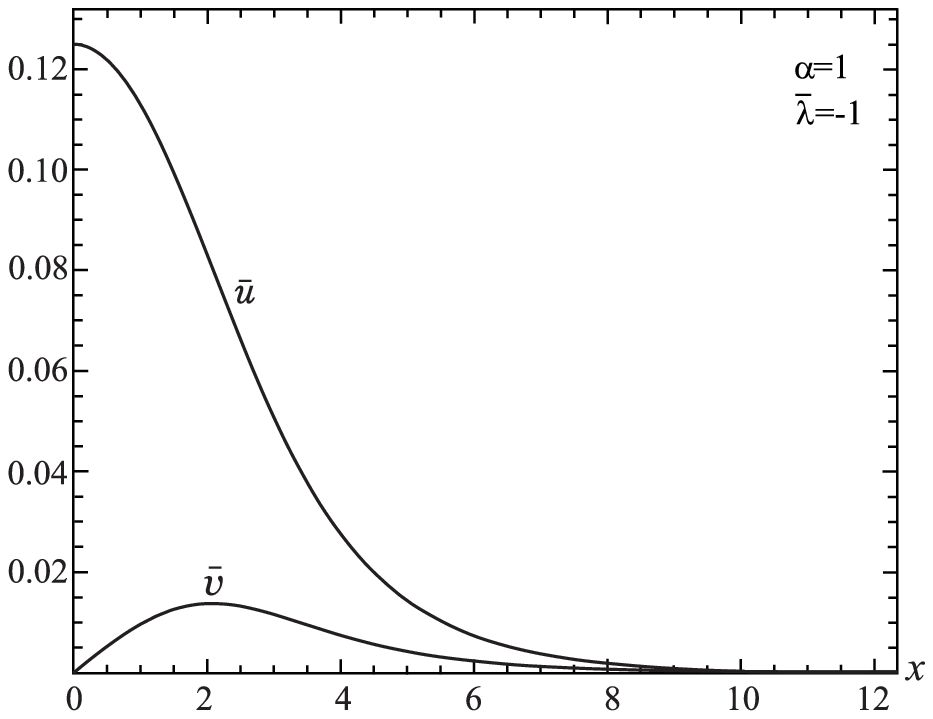}
		\end{center}
\vspace{-0.5cm}
		\caption{
		Spinor fields $\bar u$ and $\bar v$  as functions of dimensionless radius $x$ for $\beta=1$ and $\beta=0.1$.
For both cases the curves are practically indistinguishable from one another.
		}
		\label{fig_field_distr_P}
	\end{minipage}\hfill
\begin{minipage}[t]{.49\linewidth}
		\begin{center}
			\includegraphics[width=1\linewidth]{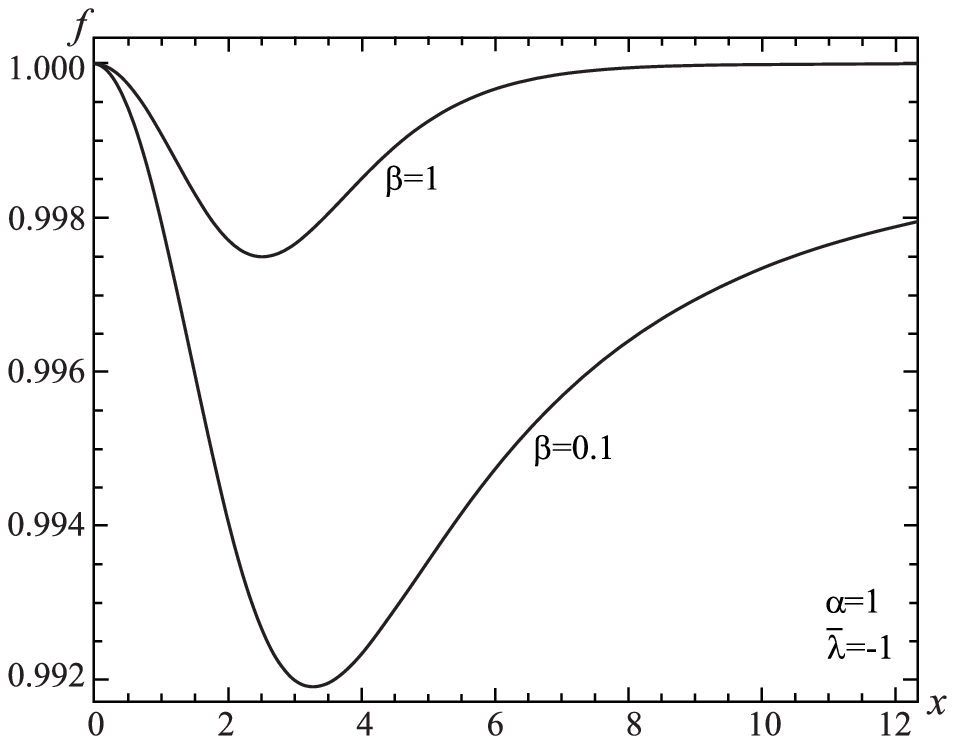}
		\end{center}
\vspace{-0.5cm}
		\caption{
		Proca field $f$ as a function of dimensionless radius $x$. Asymptotically,
both curves approach unity according to Eq.~\eqref{asymp_Proca}.
 		}
		\label{fig_field_distr_P_f}
	\end{minipage}
\end{figure}

For the Proca field, Eqs.~\eqref{fieldeqs-1_dmls}-\eqref{fieldeqs-5_dmls} are solved
when $\beta\neq 0$ and with boundary conditions assigned at the center in the form of Eq.~\eqref{bound_cond}.
The procedure for determining solutions is analogous to finding solutions in the case of the Yang-Mills field (see Sec.~\ref{YM_case}):
the interior and exterior solutions are smoothly matched on the boundary $x=x_b$. In turn,
the asymptotic solutions, obtained as $x\to \infty$, are now
\begin{equation}
\bar v\approx \bar v_\infty e^{-\sqrt{1-\bar E^2}\,x}, \quad \bar u\approx \bar u_\infty e^{-\sqrt{1-\bar E^2}\,x}, \quad f\approx 1-f_\infty e^{-\beta x},
\quad \sigma\approx 1+\ldots, \quad \bar m\approx \bar m_\infty+\ldots,
\label{asymp_Proca}
\end{equation}
where $\bar v_\infty, \bar u_\infty, f_\infty$, and $\bar m_\infty$ are integration constants, and, as in the case of the Yang-Mills field, the constant
$\bar m_\infty$ plays the role of the total ADM mass. 
Since the Proca field is massive, it decays exponentially fast with a rate
determined by the magnitude of the parameter $\beta$ [see Eq.~\eqref{asymp_Proca}].

The typical distributions of the matter fields for $\beta=0.1$ and $\beta=1$ and for the central
value $\bar u_c=0.125$ (for such $\bar u_c$ the total masses of the configurations lie near the maximum) are given in Figs.~\ref{fig_field_distr_P} and \ref{fig_field_distr_P_f}.
One can see from Fig.~\ref{fig_field_distr_P_f} that, with increasing $\beta$, the field $f$ becomes increasingly concentrated inside the radius $x_b$ where the spinor fields are nonvanishing.
In turn, the amplitude of $f$ becomes increasingly smaller and it always remains of the order of unity.
Meanwhile, the behavior of the spinor fields is practically independent of~$\beta$ (see Fig.~\ref{fig_field_distr_P}). The same is true for
the total mass of the system whose dependence on $\bar u_c$ is very similar to the case of the massless vector field considered in Sec.~\ref{YM_case} (see Fig.~\ref{fig_mass_uc}).
This is because, even if $\beta$ is large, the contribution to the total energy density coming from the term
$\beta^2 \left(f-1\right)^2/x^2$ [see the right-hand side of Eq.~\eqref{fieldeqs-3_dmls}] remains small due to the closeness of the function $f$ to unity
(see Fig.~\ref{fig_field_distr_P_f}).

Consistent with this, one might expect that when $\alpha\to 0$ the behavior of the system with the Proca field will be similar to that of the systems with the massless Yang-Mills
field studied in Secs.~\ref{YM_case} and~\ref{YM_case_lim}. In fact, numerical calculations indicate that for small $\alpha$
the solutions to the field equations~\eqref{fieldeqs-1_dmls}-\eqref{fieldeqs-5_dmls} for the spinor fields with
 $\beta\neq 0$ and $\beta= 0$ practically coincide. In turn, the Proca field always gives only a small contribution to the energy density and to the total mass of the systems.
 The same is true for the effective radius of the limiting configurations which are obtained as $\alpha\to 0$: it will be given by the expression~\eqref{R_eff_alpha_small}.

\section{Conclusions and discussion}
\label{concl}

In previous papers we considered compact, strongly gravitating configurations consisting either of only a nonlinear spinor field~\cite{Dzhunushaliev:2018jhj} or
a nonlinear spinor field together with Abelian Maxwell and Proca fields~\cite{Dzhunushaliev:2019kiy}.
In the present paper, we have studied objects supported by a nonlinear spinor field and non-Abelian SU(2) Yang-Mills/Proca fields.
To model the spinor field, we have applied the two-field harmonic {\it Ansatz}~\eqref{spinor};
this allows us to ensure that the objects under investigation are  spherically symmetric and static.
In turn, to model the SU(2) Yang-Mills/Proca fields, we have employed the standard monopole {\it Ansatz} given by Eqs.~\eqref{2-20} and \eqref{2-13},
which describes a radial magnetic field. For such a system, we have found regular spherically symmetric and asymptotically flat solutions.
We showed that such solutions can describe configurations possessing a positive ADM mass,
whose magnitude is basically determined by the central value of the spinor field and its mass $\mu$
(or, equivalently, by the value of the parameter $\alpha$).

As in the case of the systems of Refs.~\cite{Dzhunushaliev:2018jhj,Dzhunushaliev:2019kiy},
our purpose here is to obtain objects with masses of the order of the Chandrasekhar mass. To do this, we have considered the cases with different
values of the parameter $\alpha$. The results obtained can be summarized as follows.
\begin{enumerate}
\itemsep=-0.2pt
\item[(I)] For the case of the Yang-Mills field, the families of equilibrium
configurations can be parametrized by one parameter $\alpha$, whose magnitude determines the value of the maximum total ADM mass
of the configurations under consideration.  We showed that the behavior of the Yang-Mills field depends strongly on the value of $\alpha$: there is some critical $\alpha_{\text{crit}}$ separating
topological (kinklike) solutions (existing for $\alpha<\alpha_{\text{crit}}$) and nontopological solutions (existing for $\alpha>\alpha_{\text{crit}}$).

In the limit $\alpha\to 0$, the complete set of equations can be replaced by the approximate equations
that do not already involve $\alpha$ explicitly. Then the dependence of the maximum mass on
  $\alpha$  [see Eq.~\eqref{M_max_approx_2}] can be represented as
$$
M_*^{\text{max}} \approx 0.41\sqrt{\frac{|\bar \lambda|}{\alpha}}  \frac{M_\text{Pl}^2}{\mu}
\approx 0.19\, \sqrt{|\tilde \lambda|} \,M_{\odot}\left(\frac{\text{GeV}}{\mu}\right)^2.
$$
Taking into account that $\tilde \lambda$ is assumed to be of the order of unity (see the end of Sec.~\ref{prob_statem}),
this mass is comparable to the Chandrasekhar mass for the typical mass of a fermion $\mu\sim 1~\text{GeV}$.

In turn, the dependence of the effective radii of the limiting configurations with maximum masses on
    $\alpha$  [see Eq.~\eqref{R_eff_alpha_small}] can be represented as
$$
R_*^{\text{max}}\approx 1.08 \sqrt{\frac{|\bar \lambda|}{\alpha}}\lambda_c \approx
0.73\, \sqrt{|\tilde \lambda|}\left(\frac{\text{GeV}}{\mu}\right)^2 \,\, \text{km}.
$$
 For $\mu\sim 1~\text{GeV}$, the above expression gives radii of the order of kilometers. In combination  with the masses
of the order of the Chandrasekhar mass (see above), this corresponds to characteristics typical for neutron or boson stars~\cite{Schunck:2003kk,Liebling:2012fv,Mielke:2000mh,Mielke:2016war}.

\item[(II)] In the case of the Proca field, apart from the parameter $\alpha$, the system involves one more free parameter $\beta$ equal to the ratio of the Proca mass
to the mass of the spinor field. In the limit $\beta\to 0$, we return to the results of item (I).
 When $\beta\neq 0$, the vector field is not already a long-range one, and it decreases exponentially fast with distance according to the asymptotic law given by Eq.~\eqref{asymp_Proca}.

 Numerical calculations indicate that for the values of the parameter $\beta\sim 1$ considered in the present paper the contribution of the vector field to the energy density
 is always much smaller than that of the spinor field. As a result, the total mass of the configurations does not practically depend on the value of $\beta$
 and, as in the case of the massless Yang-Mills field, it can be parametrized using only the parameter $\alpha$.
 In this connection, the physical properties (namely, the total masses and effective radii) of the systems with the Proca field will practically coincide with those of configurations with the Yang-Mills field
 described in item (I).

Also, we emphasize that, unlike the Yang-Mills field, because of the presence of the last term in Eq.~\eqref{fieldeqs-5_dmls}, there are no kinklike solutions for the Proca field.
\end{enumerate}

Notice also the following feature of the systems under investigation. When considering the limiting configurations with $\alpha\to 0$ in Sec.~\ref{YM_case_lim},
one obtains the approximate solutions describing Dirac stars of the type studied in Ref.~\cite{Dzhunushaliev:2018jhj}, on the background of which a radial
SU(2) magnetic field is now present. Unfortunately, the approximate equations of Sec.~\ref{YM_case_lim} do not allow us to calculate the structure of such a magnetic field.
Nevertheless, in having the results of computations for $f$ at small $\alpha$ (see Fig.~\ref{fig_f_alpha}), one might expect that for
$\alpha\to 0$ the solution for the magnetic field will qualitatively behave as the Bartnik-McKinnon solution. In particular, apart from the one-node solution for $f$
obtained here (the curve for $\alpha=0.01$ in Fig.~\ref{fig_f_alpha}), there can also exist multinode solutions of the type found by Bartnik and McKinnon~\cite{Bartnik:1988am,Volkov:1998cc}.
Unfortunately, from the technical point of view, a search for exact solutions for the case of $\alpha\to 0$ is an extremely complicated problem since one must determine to high-accuracy eigenvalues
of the system parameters. Nevertheless, we hope that it will be possible in the future to achieve some progress in solving this problem.

In conclusion, a few words may be said about the question of the stability of the systems under consideration. It follows from the behavior of the mass--central density curves of the spinor field
(see Fig.~\ref{fig_mass_uc}) that for any $\alpha$ there is always a local maximum of the mass for some $\bar u_c$. Analogously to systems supported by, for example, scalar fields or
a hydrodynamical fluid, one could naively expect that a transition through such a local maximum must lead to instability against perturbations.
That is, the configurations located to the left of the maximum are stable, and those located to the right are unstable.
However, this problem must be studied separately by analyzing, for instance, the behavior of perturbations, as is done for boson stars~\cite{Gleiser:1988ih,Jetzer:1989us}
and systems supported by linear spinor fields~\cite{Finster:1998ws,Daka:2019iix},
or by using catastrophe theory~\cite{Kusmartsev:1990cr}.

\section*{Acknowledgments}
The authors gratefully acknowledge support provided by Grant No.~BR05236494
in Fundamental Research in Natural Sciences by the Ministry of Education and Science of the Republic of Kazakhstan. We are grateful to the Research Group
Linkage Programme of the Alexander von Humboldt Foundation for the support of this research
and would like to thank the Carl von Ossietzky University of Oldenburg for hospitality while this work was carried out.
We wish also to thank the anonymous referee whose comments helped improve the manuscript.

\end{document}